\documentclass{aa}

\usepackage{graphics}

\begin{document}
\thesaurus{05(02.01.2; 07.19.1; 08.06.2; 08.16.2; 10.07.2; 10.15.1)}

\title{Suppression of giant planet formation in stellar clusters}
	
\author{Philip J. Armitage}
\institute{Max-Planck-Institut f\"ur Astrophysik, 
	Karl-Schwarzschild-Str. 1,
        D-85741 Garching, Germany \\
	email: {\tt armitage@mpa-garching.mpg.de}} 

\date{Received / Accepted}	
	
\maketitle

\begin{abstract}
Photoevaporation driven by the ultraviolet radiation 
from massive stars severely limits the lifetime 
of protoplanetary discs around stars formed within stellar clusters. 
I investigate the resulting influence of clustered 
environments on the probability of giant planet formation, and 
show that for clusters as rich, or richer than, Orion, the time 
available for planet formation is likely to be limited to the length of any 
delay between low mass and high mass star formation. 
Under popular models for the formation of massive planets, 
the fraction of stars with giant planets in rich clusters 
is expected to be substantially suppressed as compared to 
less clustered star formation environments.
\end{abstract}

\section{Introduction}
Almost all low mass stars are born in binary systems 
(Reipurth \& Zinnecker \cite{reipurth}; Ghez et al. \cite{ghez}; 
Kohler \& Leinert \cite{kohler}), which are 
themselves often part of larger stellar clusters (Clarke, Bonnell \& 
Hillenbrand \cite{clarke}). In some 
well-studied star forming regions, such as Taurus-Auriga, these are 
merely loose aggregates of young stars (Gomez et al. \cite{gomez}), 
which may be expected to have little influence on the evolution 
of protoplanetary discs. This is not, however, the case 
in richer clusters -- of which Orion is the nearby prototype -- where 
the radiation and winds from massive stars have a dramatic impact 
on the local environment (Palla \& Stahler \cite{palla}; 
Johnstone \& Bertoldi \cite{johnstone1}; Chevalier \cite{chevalier}). Planet 
formation in the discs around low mass stars in such 
clusters cannot then be considered in isolation.

Observations of discs in Orion suggest that photoevaporation 
in the radiation field of the massive stars is likely to be 
the dominant process leading to the destruction of discs 
(Johnstone, Hollenbach \& Bally \cite{johnstone2}). In this 
paper, I assume that this is the case, and discuss the implications 
for planet formation in clusters of varying richness.
The basic conclusion is that the formation of giant planets 
in clusters richer than Orion is likely to be heavily 
suppressed, {\em unless} either they form from prompt hydrodynamic 
collapse (e.g. Boss \cite{boss}), or there is a substantial 
delay between the onset of low mass and high mass star 
formation. Preliminary indications from a {\em Hubble Space 
Telescope} search for planetary transits in the globular 
cluster 47 Tucanae (Gilliland et al. \cite{gilliland}; Brown 
et al. \cite{brown}), in which the fraction of giant short-period 
planets appears to be at least an order of magnitude below 
the value in the solar neighbourhood (Marcy \& Butler \cite{marcy}), 
are consistent with this conclusion (provided, of course, that we assume 
that planet-bearing stars in the solar neighbourhood were 
{\em not} themselves formed within rich clusters). With hindsight, 
however, numerous plausible explanations for the deficit 
of planets in this dense, low metallicity system are likely to 
be forthcoming (see e.g. Bonnell et al. 2000).

\section{Disc lifetime in stellar clusters}

\subsection{Ultraviolet flux}
Theoretical estimates suggest that there is no reason why 
protoplanetary discs around isolated low mass stars should not
survive for lengthy periods, especially 
if the rate of angular momentum transport (and hence accretion) within the disc is 
reduced due to the low ionization fraction at a few AU (Matsumoto \& Tajima \cite{matsumoto}; 
Gammie \cite{gammie}; Armitage, Livio \& Pringle \cite{armitage}). Although 
estimates of the ages of pre-main-sequence stars are subject to 
significant uncertainties (Tout, Livio \& Bonnell \cite{tout}), some 
low mass stars ($M \approx 0.5 M_\odot$) do appear to retain their 
discs for more than 10 Myr (Strom \cite{strom}; Brandner et al. \cite{brandner}). 
The disc lifetime 
is found to be much shorter for more massive stars (Strom \cite{strom}), 
and this, coupled with the observation of substantial mass loss rates 
from discs in the Orion Nebula (McCullough et al. \cite{mccullough}; 
Johnstone, Hollenbach \& Bally \cite{johnstone2}; and references 
therein), suggests that photoevaporation of the discs is an important 
process that can lead to their destruction (for a 
review see e.g. Hollenbach, Yorke \& Johnstone \cite{hollenbach})

Photoevaporation of discs has been 
extensively studied, both when the source of ionizing radiation 
is the central star (Shu, Johnstone \& Hollenbach \cite{shu}; 
Hollenbach et al. \cite{hollenbach2}), and 
for the case relevant here where the radiation field arises externally 
(Johnstone, Hollenbach \& Bally \cite{johnstone2}; St\"orzer \& 
Hollenbach \cite{storzer}).
In the simplest analysis, ultraviolet radiation 
heats the disc surface, raising the sound speed to $c_s$. Beyond 
a radius,
\begin{equation} 
 R_{\rm g} \approx 0.5 { {GM_*} \over c_s^2 }
\label{eq1}
\end{equation} 
the heated gas is unbound and flows away as a thermal wind. 
Both Lyman continuum EUV photons 
($\lambda < 91.2 \ {\rm nm}$, $h \nu > 13.6 \ {\rm eV}$), and less 
energetic far-ultraviolet (FUV) radiation ($6 \ {\rm eV} \ < h \nu < \ 
13.6 \ {\rm eV}$) can contribute to disc mass loss. However, 
for $M_* \approx M_\odot$, FUV-driven flows (for which the 
sound speed in the heated layer is $\sim 3 {\rm kms}^{-1}$, as 
compared to $\sim 10 {\rm kms}^{-1}$ for EUV-driven flows) have $R_{\rm g}$ 
that is substantially larger than the radii of principal interest 
for planet formation. In this paper we therefore concentrate on 
EUV-driven mass loss, while noting that the effects of FUV 
radiation will also be important for lower mass stars.

To estimate the strength of ultraviolet radiation in clusters 
of varying richness, we first need the emission from an individual 
star. For stars with $0.8 \ M_\odot < M_* < 120 \ M_\odot$, we 
obtain the luminosity and effective temperature $T_e$ 
from stellar models computed by Schaller et al. (\cite{schaller}) 
for a metallicity $Z=0.02$ (for these purposes, differences in 
metallicity are of secondary importance). We use the model output 
closest to $t = 10^6 \ {\rm yr}$, which should be appropriate for 
young clusters of any age given the very short pre-main-sequence 
phase of the massive stars that dominate the UV flux. The model 
output is then combined with theoretical stellar atmosphere models 
by Kurucz (e.g. Buser \& Kurucz \cite{buser}) to yield the EUV 
and FUV output as a function of stellar mass. 

\begin{figure}[t]
 \resizebox{\hsize}{!}{\includegraphics{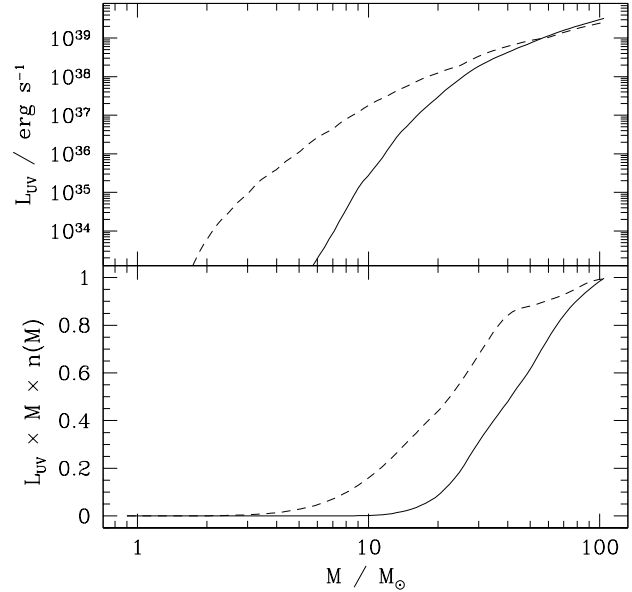}}
 \vspace{-0.3truein} 
 \caption{Upper panel, the luminosity in EUV (solid lines) and 
          FUV (dashed lines) for stars of mass $M_*$. The lower 
	  panel shows, for a Salpeter mass function, the expected 
	  relative contribution to the total flux from stars of different 
	  masses. The units in the lower panel are arbitrary.}	  
 \label{fig1}
\end{figure}

Fig.~{\ref{fig1}} shows the EUV and FUV luminosity as a function of stellar 
mass. A negligible fraction of the total flux is emitted in the EUV by stars with 
$T_e < 2.5 \times 10^4 \ {\rm K}$, which corresponds to a mass 
$M_* \approx 12 M_\odot$. The fraction of flux in the EUV band 
then rises roughly linearly with increasing $T_e$ to $\sim 0.6$ of the total for 
$M_* \sim 10^2 M_\odot$ stars with $T_e \approx 5 \times 10^4 \ {\rm K}$. 
Significant fluxes of FUV radiation are produced by lower 
mass stars with $T_e > 10^4 \ {\rm K}$.

For a standard mass function the most massive stars are rare.
Fig.~{\ref{fig1}} also shows which stars dominate the output of EUV and FUV 
radiation in a cluster. We assume a Kroupa, Tout \& Gilmore (\cite{kroupa}) 
mass function for $M_* < 1 \ M_\odot$, and a Salpeter (\cite{salpeter}) 
form, $n(M) dM \propto M^{-2.35}$, for higher masses. Both the FUV and 
(especially) the EUV flux is expected to be dominated by the most massive 
star in a cluster. 

\begin{figure}[t]
 \resizebox{\hsize}{!}{\includegraphics{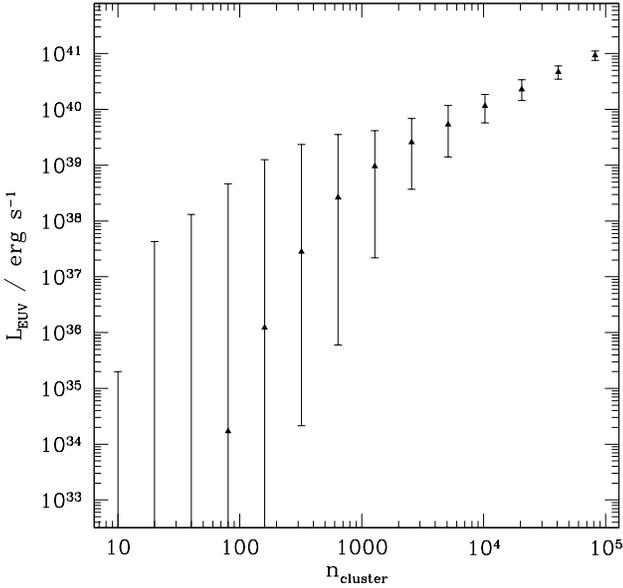}}
 \vspace{-0.3truein} 
 \caption{The integrated EUV luminosity for clusters with 
          $n_{\rm cluster}$ stars. The points show median
	  values, the error bars show the range that encloses 
	  90\% of the distribution of luminosities.}	  	  
 \label{fig2}
\end{figure}

Using this mass function, with a low mass cutoff at 
$0.1 \ M_\odot$, and a high mass cutoff at $120 \ M_\odot$, we generated 
realizations of clusters with number of members $n_{\rm cluster}$ ranging 
from $10$ to $10^5$. Fig.~{\ref{fig2}} shows the resultant integrated EUV 
luminosities. The median luminosity rises steeply for smaller clusters, which 
do not fully sample the high mass end of the mass function, and becomes 
linear only for larger clusters with $n_{\rm cluster} > 10^3$. The 
dependence of the luminosity on the single most massive star that a cluster 
happens to have implies a large dispersion in the integrated luminosity. 
The range of luminosities that encompasses 90\% of the probability 
distribution spans almost 3 orders of magnitude at $n_{\rm cluster} = 10^3$, 
and is still a factor of $\sim 3$ at $n_{\rm cluster} = 10^4$. This 
implies that otherwise similar clusters in which the maximum stellar mass 
varies can have widely different UV environments, leading to substantial 
dispersion in the predicted lifetimes of circumstellar discs.

\subsection{Disc lifetime}

The theory of EUV-driven flows from discs provides a prediction for 
how the mass loss rate scales with the flux of ionizing photons. 
In the simplest analysis, the mass loss rate for a disc of 
fixed radius, at distance $d$ 
from a source emitting a flux of ionizing photons $\Phi$, scales as 
(Bertoldi \& McKee \cite{bertoldi}; Johnstone, Hollenbach \& 
Bally \cite{johnstone2}),
\begin{equation} 
 \dot{M}_{\rm outflow} \propto \Phi^{1/2} d^{-1}.
 \label{eq2}
\end{equation}  
More sophisticated analysis are available (St\"orzer \& Hollenbach 
\cite{storzer}; Richling \& Yorke \cite{richling}), but are hardly 
warranted here given the uncertainties.

To calibrate the predicted disc lifetime, we make use of the results 
obtained in Orion. The most massive star in the Orion Trapezium, 
$\theta^1$ Ori C, has a mass estimated at around $40-50 \ M_\odot$  
(Hillenbrand \cite{hillenbrand}). Both this mass, and the estimated 
ionizing flux in Orion of the order of $\Phi \sim 10^{49} \ {\rm s}^{-1}$, 
fall withing the range expected for clusters with $n \sim 10^3$. 
Within Orion, one of the best studied evaporating 
discs, HST 182--413, at a projected distance of 0.12 pc from $\theta^1$ Ori C, 
has an estimated mass loss rate 
$\dot{M}_{\rm outflow} = 4.1 \times 10^{-7} \ M_\odot {\rm yr}^{-1}$. 
The disc radius $r_{\rm disc}$ is $\approx 100 \ {\rm AU}$, and the estimated disc 
mass $0.04 \ M_\odot$ (Johnstone, Hollenbach \& 
Bally \cite{johnstone2}). This disc mass is consistent with the 
upper end of the distribution of disc masses inferred from 
mm-wavelength observations by 
Osterloh \& Beckwith (\cite{osterloh}). For this object, the 
characteristic disc lifetime is therefore
$t_{\rm disc} \equiv M_{\rm initial} / \dot{M}_{\rm outflow} \approx 
10^5 \ {\rm yr}$. 

For planet formation, we are interested in smaller discs than 
that which presently surrounds HST 182--413. These will have smaller 
mass loss rates, which in the case of EUV-dominated flows scale as 
$r_{\rm disc}^{3/2}$ (Johnstone, Hollenbach \& 
Bally \cite{johnstone2}). Scaling to a solar system sized disc (30 AU), 
and assuming conservatively that the mass remains similar to that 
of HST 182--413, we obtain $t_{\rm disc} \simeq 6 \times 10^5 \ {\rm yr}$ as a simple 
estimator of the disc survival time for $\Phi = 10^{49} \ {\rm s}^{-1}$ 
and $d=0.12 \ {\rm pc}$. This estimate of the time available 
for planet formation before the disc is destroyed is certainly crude, and in 
particular ignores completely the effects of angular momentum transport 
in the disc. However, viscosity will tend to {\em hasten} disc destruction by moving 
material to larger radii where it can be lost via photoevaporation more 
easily. 

\begin{figure} 
 \resizebox{\hsize}{!}{\includegraphics{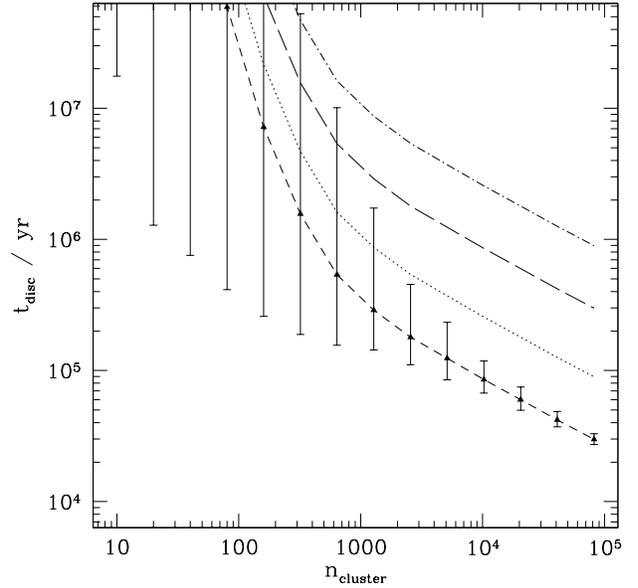}}
 \vspace{-0.3truein} 
 \caption{The predicted dependence of the disc lifetime on 
          cluster size, for discs at different radii, 0.1 pc 
	  (lower dashed line), 0.3 pc (dotted line), 1 pc 
	  (long dashed line), and 3 pc (dot-dashed line). 
	  The error bars plotted for the 0.1 pc curve show the 
	  range of predictions that include 90\% of the 
	  probability distribution. Errors on the other curves are 
	  identical.}
 \label{fig3}
\end{figure}

Fig.~{\ref{fig3}} shows the dependence of the predicted disc lifetime 
on the size of the cluster, for discs located at distances $d$ from 
the principal source of ionizing radiation. For smaller clusters 
this will typically be a single massive star, while for rich 
clusters with $n_{\rm cluster} \sim 10^4$ or larger several 
stars will contribute. We assume that, as in Orion, these 
stars are formed near the cluster centre (Bonnell \& Davies \cite{bonnell}), 
and thus can be treated as a single radiation source. If in reality 
the massive stars were instead distributed throughout the cluster, 
that would result in a {\em larger} volume over which significant 
ablation of discs would occur. Similarly, mixing of the cluster 
over timescales shorter than the disc lifetime increases the 
fraction of stars whose discs are affected by photoevaporation.

By construction, the dependence shown in Fig.~{\ref{fig3}} is 
consistent with a destruction timescale of a few $\times 10^5$ yr 
in the inner ($\sim 0.1$ pc) part of Orion near $\theta^1$ Ori C. 
It is also consistent with observed disc lifetimes of a few Myr, extending 
upwards to more than $10^7$ yr, in poor clusters with 
just a few hundred members. A large dispersion in disc lifetime 
is expected in this case. 

For clusters richer than Orion, the volume over which photoevaporation 
significantly curtails the disc lifetime grows. For large 
$n_{\rm cluster}$, $\Phi \propto n_{\rm cluster}$, and the volume 
within which mass loss is significant scales as $n_{\rm cluster}^{3/2}$. 
Thus, for rich clusters, a larger fraction of the discs will be 
ablated by the action of the radiation field, even if the central 
cluster density remains constant. For $n_{\rm cluster} = 10^5$, 
the disc lifetime is expected to be severely curtailed (less than 
a few $\times 10^5$ yr) out 
to distances of the order of 1 pc, and to be significantly 
reduced (to $< 10^6$ yr, an order of magnitude below the value 
inferred for isolated discs), out to distances of perhaps 
3 -- 5 pc. 

\subsection{Planet formation}

In the Solar System, the observation that the gas fraction 
of the outer planets varies substantially has been taken 
to imply that these planets formed at roughly the epoch 
when the gas disc was being dissipated, probably at 
about a few $\times 10^6 - 10^7$ yr (e.g. Shu, Johnstone \& 
Hollenbach \cite{shu}; St\"orzer \& Hollenbach \cite{storzer}).
A few Myr is also the 
typical theoretical estimate for the time required 
for a growing core at 5 AU to reach a mass where it can rapidly 
accrete gas from the disc in a runaway manner (Pollack et al. 
\cite{pollack}). Somewhat shorter timescales are derived 
if the cores initially migrate inwards through the gas disc, 
but subsequently halt and accrete the gas at smaller radii 
(Papaloizou \& Terquem \cite{papaloizou}). These timescales 
suggest that a disc lifetime of $10^6$ yr might well pose 
difficulties for giant planet formation, at least in some 
theoretical models, while lifetimes closer to $10^5$ yr 
would very probably preclude giant planet formation. From 
Fig.~{\ref{fig3}}, this suggests that giant planet formation 
will be strongly suppressed in clusters with $\sim 10^5$ 
stars out to around 1 pc, and possibly at substantially 
larger radii. We also note that since 
short period planets (analagous to that orbiting 51 Peg) 
are unlikely to have been formed {\em in situ} 
(Bodenheimer, Hubickyj \& Lissauer \cite{bodenheimer}), 
the disc must survive for a significant additional time, 
subsequent to planet formation, to allow inward migration through 
the gas disc (Lin, Bodenheimer \& Richardson \cite{lin}). Some 
additional reduction in the frequency of such systems would 
then result. Low mass planets are largely unaffected by these 
considerations. Indeed, since rapid inward migration through 
the disc can substantially deplete the population of such 
objects (e.g. Ward \cite{ward}, and references therein), the 
rapid removal of the gas could in principle even 
enhance the initial frequency of low mass planets in clusters.  

The conclusion that rapid disc destruction in clusters severely 
reduces the probability of planet formation can be evaded 
in two ways. First, massive planets might form from the disc 
via direct hydrodynamic collapse (e.g. Boss \cite{boss}, \cite{boss2}; 
Cameron \cite{cameron}). 
If this process occurs at all, it is most likely at very 
early times when the disc is massive and vulnerable to 
gravitational instability. A detection of massive planets 
around systems where the disc lifetime was very short 
would constitute indirect but persuasive evidence for the importance 
of this process. Second, if there is a significant delay 
between the epoch of low mass and high mass star formation, 
planets could have time to form before the discs began 
to be exposed to ionizing radiation. In Orion, observations 
suggest that at least 
some low mass star formation appears to have been 
underway well before the formation of the high mass 
members of the cluster (Palla \& Stahler \cite{palla}),  
while in some theoretical models for high mass star formation 
prior low mass star formation is a necessary ingredient 
(Bonnell, Bate \& Zinnecker \cite{bonnell2}). More 
generally, unless star formation in a cluster is somehow synchronized 
to better than the sound crossing time of the gas in the 
star forming region, there is bound to be some spread 
in the times at which stars form. Since high mass 
stars are relatively scarce, this would typically lead to 
some low mass stars in the cluster forming well before the first massive 
star turns on and begins the process of disc destruction.
 
\section{Discussion}
As searches for extrasolar planets extend beyond the immediate 
solar neighbourhood, it will become possible to study how the 
frequency and properties of extrasolar planetary systems depend 
upon the star formation environment. In this paper, I have 
argued that in what may be the typical setting for star formation 
(Clarke, Bonnell \& Hillenbrand \cite{clarke}) -- a 
cluster containing a mix of high and low mass stars -- photoevaporation 
of protoplanetary discs severely limits the time available for giant
planet formation. Observations indicate that discs are being rapidly 
destroyed by this process within a few tenths of a pc of Orion's 
Trapezium, and the effect will be stronger in richer clusters. 
For a cluster of $10^5$ stars, the disc lifetime is likely to be 
significantly reduced when compared to that of discs around isolated stars,
out to a distance from the cluster centre of several pc. Unless 
planets form contemporaneously with the disc itself (Boss \cite{boss}), 
or there is a significant delay (perhaps $10^6$ yr or more) between the 
epochs of low mass and high mass star formation, the fraction of stars 
in rich clusters with massive planets is likely to be small. This 
provides both a possible explanation for the apparent dearth of massive, 
short-period planets in 47 Tucanae (Gilliland et al. \cite{gilliland}; Brown 
et al. \cite{brown}), and suggests that planet formation could also 
be noticeably suppressed in substantially less dense clustered environments.

\acknowledgements
I thank Melvyn Davies for useful discussions, 
and the referee for a prompt and helpful report.


\begin{thebibliography}{}

\bibitem[2000]{armitage} 
 Armitage, P.J., Livio, M., Pringle, J.E., 2000, MNRAS, submitted

\bibitem[1990]{bertoldi} 
 Bertoldi, F., McKee, C.F., 1990, ApJ, 354, 529
 
\bibitem[2000]{bodenheimer}
 Bodenheimer, P., Hubickyj, O., Lissauer, J.J., 2000, Icarus, 143, 2 
 
\bibitem[1998]{bonnell2}
 Bonnell, I.A, Bate, M.R., Zinnecker, H., 1998, 
 MNRAS, 298, 93
 
\bibitem[1998]{bonnell}
 Bonnell, I.A., Davies, M.B., 1998, MNRAS, 295, 691
 
\bibitem[2000]{bonnell3}
 Bonnell, I.A., Smith, K.W., Davies, M.B., Horne, K., 2000, MNRAS, submitted  

\bibitem[1998]{boss}
 Boss, A.P., 1998, 503, 923
 
\bibitem[2000]{boss2}
 Boss, A.P., 2000, ApJ, 536, L101 
 
\bibitem[2000]{brandner}
 Brandner, W., et al., 2000, AJ, 120, 950
 
\bibitem[2000]{brown}
 Brown, T.M., et al., 2000, American Astronomical Society Meeting 196, \#02.03
 
\bibitem[1992]{buser} 
 Buser, R., Kurucz, R.L., 1992, A\&A, 264, 557 
 
\bibitem[1978]{cameron} 
 Cameron, A.G.W., 1978, Moon Planets, 18, 5  
 
\bibitem[2000]{chevalier}
 Chevalier, R.A., 2000, ApJ, 538, L151
 
\bibitem[2000]{clarke}
 Clarke, C.J., Bonnell, I.A., Hillenbrand, L.A., 2000, in 
 Protostars and Planets IV, eds V. Mannings, A.P. Boss, and S.S. Russell, 
 University of Arizona press, p.~151 
 
\bibitem[1996]{gammie}
 Gammie, C.A., 1996, ApJ, 457, 355 

\bibitem[1997]{ghez}
 Ghez, A.M., McCarthy, D.W., Patience, J.L., Beck, T.L., 1997, 
 ApJ, 481, 378 
 
\bibitem[2000]{gilliland}
 Gilliland, R.L., et al., 2000, American Astronomical Society Meeting 196, \#02.02
 
\bibitem[1993]{gomez}
 Gomez, M., Hartmann, L., Kenyon, S.J., Hewett, R., 1993, 
 AJ, 105, 1927 
 
\bibitem[1997]{hillenbrand}
 Hillenbrand, L.A., 1997, AJ, 113, 1733 
 
\bibitem[1994]{hollenbach2}
 Hollenbach, D., Johnstone, D., Lizano, S.,
 Shu, F., 1994, ApJ, 428, 654 
 
\bibitem[2000]{hollenbach}
 Hollenbach, D.J., Yorke, H.W., Johnstone, D., 2000, in 
 Protostars and Planets IV, eds V. Mannings, A.P. Boss, and S.S. Russell, 
 University of Arizona press, p.~401  
 
\bibitem[2000]{johnstone1}
 Johnstone, D., Bertoldi, F., 2000, in The Orion Complex 
 Revisited, eds  M.J. McCaughrean \& A. Burkert, ASP Conf. Ser., 
 in press
 
\bibitem[1998]{johnstone2}
 Johnstone, D., Hollenbach, D., Bally, J., 1998, ApJ, 499, 758 
 
\bibitem[1998]{kohler}
 Kohler, R., Leinert, C., 1998, A\&A, 331, 977
 
\bibitem[1990]{kroupa}
 Kroupa, P., Tout, C.A., Gilmore, G., 1990, MNRAS, 244, 76 
 
\bibitem[1996]{lin}
 Lin, D.N.C., Bodenheimer, P., Richardson, D.C., 1996, 
 Nature, 380, 606 
 
\bibitem[2000]{marcy}
 Marcy, G.W., Butler, R.P., 2000, PASP, 112, 137 
 
\bibitem[1995]{matsumoto}
 Matsumoto, R., Tajima, T., 1995, ApJ, 445, 767 
 
\bibitem[1995]{mccullough}
 McCullough, P.R., et al., 1995, ApJ, 438, 394 
 
\bibitem[1995]{osterloh} 
 Osterloh, M., Beckwith, S.V.W., 1995, ApJ, 439, 228 
 
\bibitem[1999]{palla} 
 Palla, F., Stahler, S.W., 1999, ApJ, 525, 772 
 
\bibitem[1999]{papaloizou}
 Papaloizou, J.C.B., Terquem, C., 1999, ApJ, 521, 823 
 
\bibitem[1996]{pollack}
 Pollack, J.B., et al., 1996, Icarus, 124, 62  
 
\bibitem[1993]{reipurth} 
 Reipurth, B., Zinnecker, H., 1993, A\&A, 278, 81 
 
\bibitem[1998]{richling} 
 Richling, S., Yorke, H.W., 1998, A\&A, 340, 508 
 
\bibitem[1955]{salpeter}
 Salpeter, E.E., 1955, ApJ, 121, 161 
 
\bibitem[1992]{schaller}
 Schaller, G., Schaerer, D., Meynet, G., Maeder, A., 1992, A\&AS, 96, 269
 
\bibitem[1993]{shu}
 Shu, F.H., Johnstone, D., Hollenbach, D., 1993, Icarus, 106, 92 
 
\bibitem[1999]{storzer}
 St\"orzer, H., Hollenbach, D., 1999, ApJ, 515, 669 
 
\bibitem[1995]{strom}
 Strom, S.E., 1995, Rev. Mex. Astron. Astrophys. Conf. Ser., 1, 317 
 
\bibitem[1999]{tout}
 Tout, C.A., Livio, M., Bonnell, I.A., 1999, MNRAS, 310, 360 
 
\bibitem[1997]{ward}
 Ward, W.R., 1997, Icarus, 126, 261 
 
\end{thebibliography}
\end{document}